\documentclass[8pt,preprintnumbers,amsmath,amssymb,prl,twocolumn,showpacs]{revtex4}
\usepackage{graphicx}
\usepackage{dcolumn}
\usepackage{bm}
\usepackage[colorlinks=true,linkcolor=red,anchorcolor=red,citecolor=red,urlcolor=blue]{hyperref}

\begin{document}

\title{Quantum Transport in Magnetic Topological Insulator Thin Films}

\author{Hai-Zhou Lu, An Zhao, and Shun-Qing Shen}

\affiliation{Department of Physics, The University of Hong Kong, Pokfulam Road, Hong Kong, China}

\date{20 June, 2013}
\begin{abstract}
The experimental observation of the long-sought quantum anomalous
Hall effect was recently reported in magnetically doped topological
insulator thin films [Chang \emph{et al}., Science \textbf{340}, 167 (2013)].
An intriguing observation is a rapid decrease from the quantized plateau in the Hall conductance, accompanied by a peak in
the longitudinal conductance as a function of the gate voltage.
Here, we present a quantum transport theory with an effective
model for magnetic topological insulator thin films. The good agreement between theory and experiment reveals that the measured transport originates from a topologically nontrivial conduction band which, near its band edge, has concentrated Berry curvature and a local maximum in group
velocity.  The indispensable roles of the broken structure inversion and particle-hole symmetries are also revealed. The results are instructive for future
experiments and transport studies based on first-principles calculations.
\end{abstract}

\pacs{73.50.-h, 73.63.-b, 85.70.-w}

\maketitle
In some metallic ferromagnets, a transverse current can be induced
by a longitudinal electric field, known as the anomalous Hall effect \cite{Nagaosa10rmp,Xiao11rmp}. The phenomenon does not need an
external magnetic field, thus it is distinct from the ordinary Hall effect.
It has been perceived that in some insulating ferromagnets the anomalous Hall conductance could be quantized in units of the conductance quantum $e^{2}/h$, meanwhile, the longitudinal conductance vanishes
\cite{Haldane-88prl}, leading to the quantum anomalous Hall effect,
the last and long-sought family member of the Hall effects.
In the quantum anomalous Hall system, the nontrivial topology of the bulk states and broken time-reversal symmetry give rise to
chiral edge states in the energy gap.
The dissipationless transport of the topologically protected edge states gives the quantized conductances, and is believed to have promising applications in quantum electronic devices with low power consumption.
Solutions and mechanisms to realize the quantum anomalous Hall effect
have attracted tremendous efforts in past decades \cite{Qi06prb,Onoda06prl,Liu08prl,Yu10sci,Nomura11prl,Qiao10rc,Chu11prb,Chang13am,Checkelsky12natphys}.
One of the most promising schemes \cite{Yu10sci} is based on the
magnetically doped topological insulators \cite{Hor10prb,Chen10sci,Wray11natphys},
where the interplay of strong spin-orbit coupling and magnetic exchange
interaction gives rise to the band inversion required by the quantum
anomalous Hall effect.

\begin{figure}[htbp]
\centering \includegraphics[width=8.5cm]{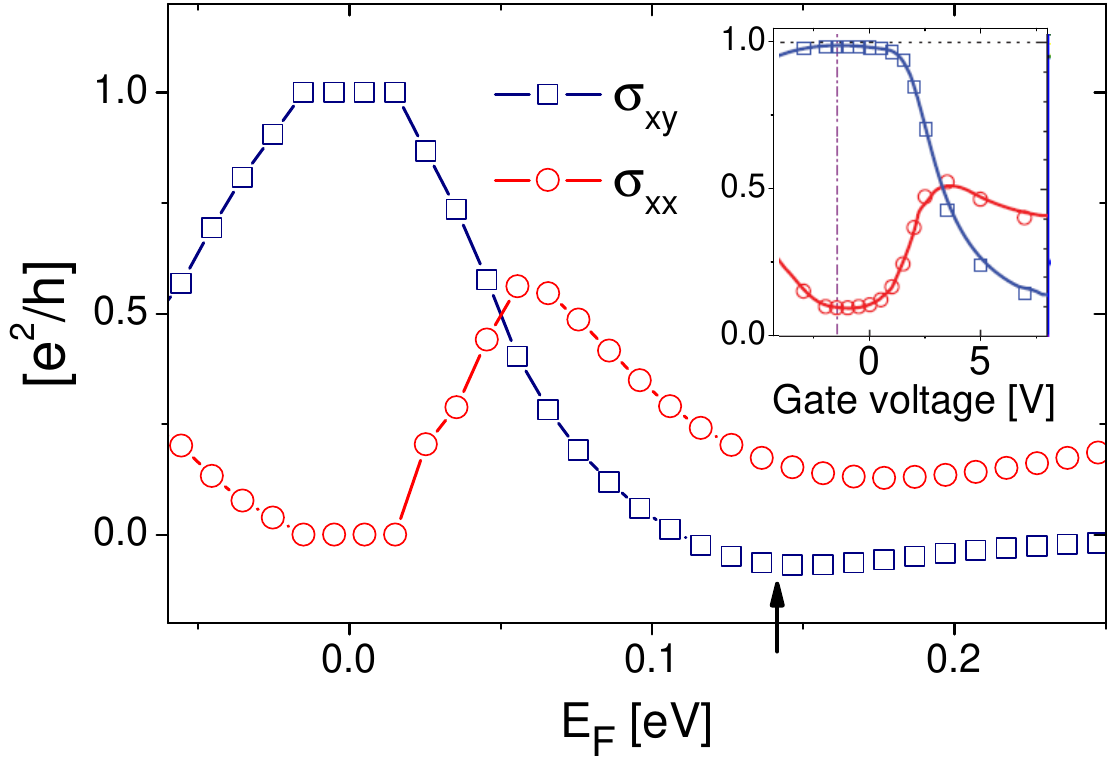} \caption{(color online). Calculated Hall (square) and longitudinal (circle) conductances as
functions of the Fermi energy $E_{F}$. Parameters: $\Delta=0.01$
eV, $B=-30$ eV$\cdot$\AA$^{2}$, $\gamma=3$ eV$\cdot$\AA, $m=-0.06$ eV,
$D=28$ eV$\cdot$\AA$^{2}$, $V=0.01$ eV, $nu_{1}^{2}=nu_{2}^{2}=100$
(eV$\cdot$\AA)$^{2}$, and $nu_{3}^{2}=nu_{4}^{2}=10$ (eV.\AA)$^{2}$.
The energy-related parameters are up to a scaling compared to the
experiment. Inset: the Hall and longitudinal conductances measured
in the experiment (adopted from Ref. \cite{Chang13sci}).}

\label{fig:sigma}
\end{figure}

Recently, the experimental observation on the quantum anomalous Hall
effect was reported in Cr-doped (Bi,Sb)$_{2}$Te$_{3}$ ultra thin
films \cite{Chang13sci}. The measured Hall conductance exhibits a
quantized plateau while the longitudinal conductance decreases drastically
at lower temperatures. A more subtle behavior appeared on the positive
gate voltage side of the quantized plateau: the Hall conductance
shows a sudden drop, accompanied by a peak in the longitudinal conductance
(the inset of Fig.\ref{fig:sigma}). Understanding the mechanisms
beneath the subtle behavior is crucial because they are closely related to the
topological origin of the quantized plateau. In this Letter, we present
a quantum transport theory, based on an effective microscopic model
for the topological insulator thin films grown on substrates. With the single model, both the calculated Hall and longitudinal conductances match the measured data very well (see Fig. \ref{fig:sigma}). The good agreement
between theory and experiment reveals the following. (1) The conduction and
valence bands closest to the quantized plateau are \emph{always} topologically
nontrivial. The transport
features in Fig. \ref{fig:sigma} come from the nontrivial conduction band, which has concentrated Berry curvature and a local maximum in group velocity near its band edge. (2) To induce
the band inversion for the quantum anomalous Hall effect, a stronger
magnetic exchange field is required to overcome not only the finite-size
gap of the thin film but also the effect of structural inversion asymmetry
(SIA), which is caused by the potential difference between the top and bottom surfaces. Usually SIA is stronger
in thicker films, resulting in a possible obstacle to realize the quantum
anomalous Hall effect in thicker samples. (3) The transport is in
the diffusive regime: the longitudinal transport depends on the group
velocity rather than the density of states, which will be helpful
for further transport studies based on first-principles calculations.
(4) The broken particle-hole symmetry is also indispensable, which gives rise to the local maximum in the group velocity and the asymmetry around the quantized
plateau.

\begin{figure}[htbp]
\centering \includegraphics[width=8.5cm]{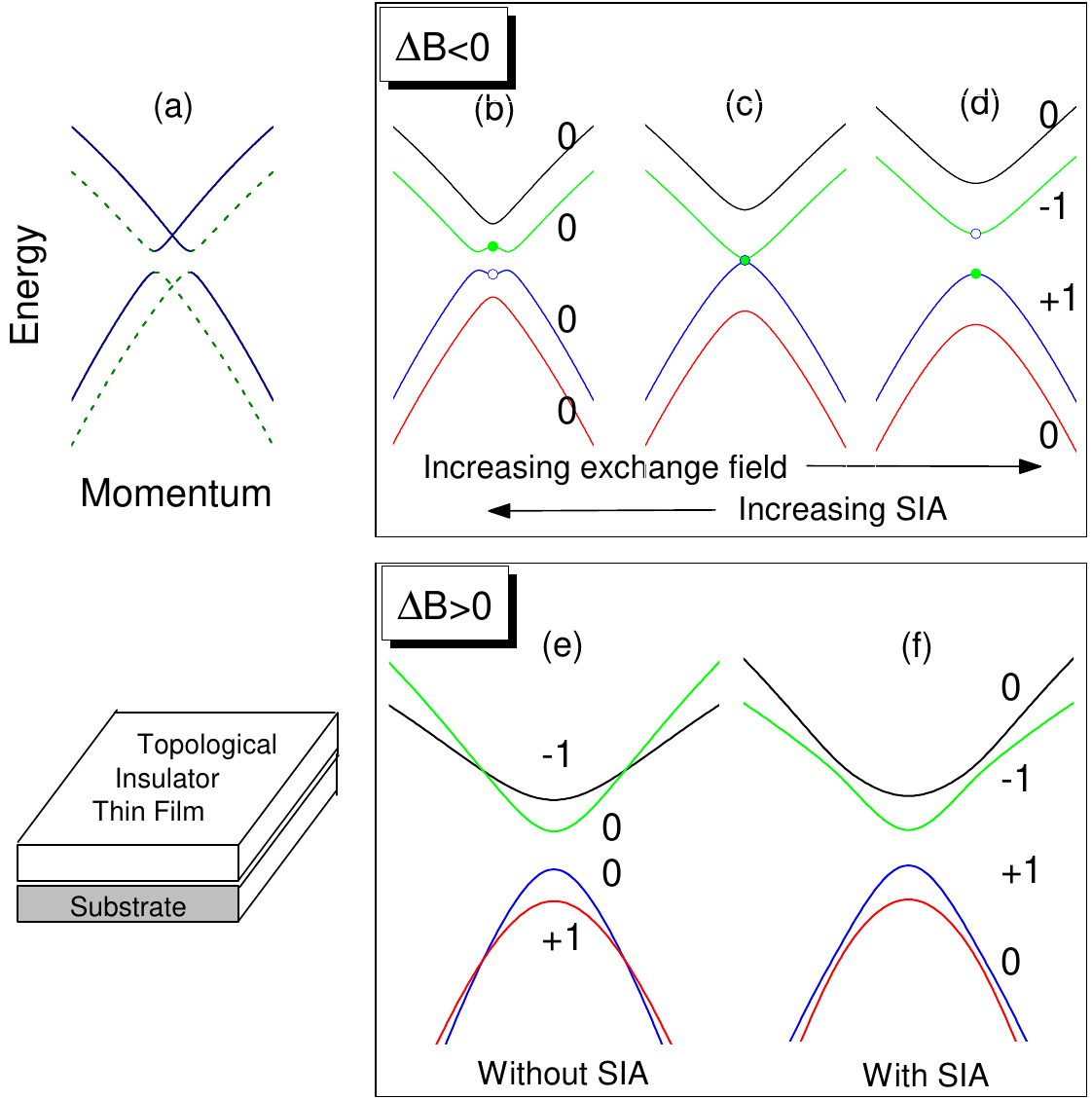} \caption{(color online). (a) Solid and dashed curves depict the top and bottom surface states,
respectively, of a thin film of topological insulator on the substrate
before magnetic doping. [(b)-(d)] The exchange field from magnetic
dopants can lift the degeneracies at $k=0$ and induce a band inversion,
whereas SIA induced by the substrate
is competing against the exchange field. (d) No band crossing at finite $k$ if $\Delta B<0$.
(e) Band crossings at finite $k$ are expected if the band inversion is achieved
by doping a thin film with $\Delta B>0$. (f) SIA can turn the band
crossings to anticrossings while exchanging the topological properties
between inner and outer bands. $0,\pm1$ indicate the contribution
to $\sigma_{xy}$ (Chern number) of a band if the band is fully occupied. }
\label{fig:evolution}
\end{figure}

Ferromagnetic insulators were predicted to form in magnetically doped
topological insulators such as Cr or Fe doped Bi$_{2}$Te$_{3}$ and
Sb$_{2}$Te$_{3}$, and their thin films have a topological
band structure, leading to the quantum anomalous Hall effect \cite{Yu10sci}.
Starting from the three-dimensional effective model for topological
insulators \cite{Zhang09np,Shen11spin} with the exchange field splitting,
and using the solution of the confined quantum states at the $\Gamma$
point ($k_{x},k_{y}=0$) in a thin film as a basis \cite{Lu10prb,Shan10njp},
we obtain the effective model \cite{supp}
\begin{eqnarray}
H=H_{0}+\frac{m}{2}\tau_{0}\otimes\sigma_{z}.\label{H}
\end{eqnarray}
$m$ is the exchange field from the magnetic dopants \cite{Liu08prl,Yu10sci},
which acts effectively like a Zeeman field. $\tau_{0}$ is a $2\times2$
unit matrix. $\sigma_{z}$ is the $z$ Pauli matrix. $H_{0}$ is the
effective model for the thin films of the topological insulator \cite{Lu10prb,Shan10njp}
\begin{eqnarray}
 &  & H_{0}=-Dk^{2}\nonumber \\
 &  & +\left(\begin{array}{cccc}
\frac{\Delta}{2}-Bk^{2} & i\gamma k_{-} & V & 0\\
-i\gamma k_{+} & -\frac{\Delta}{2}+Bk^{2} & 0 & V\\
V & 0 & -\frac{\Delta}{2}+Bk^{2} & i\gamma k_{-}\\
0 & V & -i\gamma k_{+} & \frac{\Delta}{2}-Bk^{2}
\end{array}\right)\nonumber \\
\end{eqnarray}
where $(k_{x},k_{y})$ is the wave vector, $k^{2}=k_{x}^{2}+k_{y}^{2}$.
The $D$ term breaks the particle-hole symmetry, and the band gap
opening requires $|D|<|B|$. $\Delta$ is the hybridization of the top
and bottom surface states of the thin film \cite{Lu10prb,Linder09prb,Liu10rc},
which becomes relevant for thin films, e.g., Bi$_{2}$Se$_{3}$ thinner
than 5 nm \cite{Zhang10natphys,Sakamoto10prb}. Both $\Delta$ and
$B$ are functions of the thickness of thin film, and approach zero
simultaneously for a thicker film. $\gamma=v\hbar$, with $v$ the
effective velocity. $V$ measures the SIA between the top and bottom surfaces of the thin film. The inclusion of SIA here is a natural consequence for a realistic thin film grown
on a substrate, which always induces a potential distribution along
the film growth direction \cite{Shan10njp}. The potential shifts
the gapless Dirac cones on the top and bottom surfaces, and gives
the Rashba-like splitting in the band structure when the top-bottom
hybridization $\Delta$ is present {[}Fig. \ref{fig:evolution} (a){]}
as observed by ARPES \cite{Zhang10natphys}.

It can be explicitly shown that the SIA increases the exchange field required by the quantum anomalous Hall effect. Under a unitary transformation \cite{supp}, the Hamiltonian can be
diagonalized into two $2\times2$ blocks
\begin{eqnarray*}
h_{s}= & -Dk^{2}+\sigma_{z}\left(\Gamma+s\Lambda\right)+s\gamma(k_{x}\sigma_{y}-k_{y}\sigma_{x})\cos\Theta,
\end{eqnarray*}
where $s=\pm1$ for the outer and inner blocks, respectively. The outer (inner) block has a larger (smaller) band gap at $k=0$. We have defined
$\Gamma$ = $\sqrt{(m/2)^{2}+\gamma^{2}k^{2}\sin^{2}\Theta}$, $\Lambda$
= $\sqrt{(\Delta/2-Bk^{2})^{2}+V^{2}}$, and $\cos\Theta$ = $(\Delta/2-Bk^{2})/\Lambda$.
$\sigma_{x,y,z}$ are the Pauli matrices. For $s=+1$, the dispersions
of the two bands (denoted as the outer bands) are $E_{i}=-Dk^{2}\pm\sqrt{\left(\Lambda+\Gamma\right)^{2}+(\gamma k)^{2}\cos^{2}\Theta}$
($i=1$ for $-$, and 4 for $+$), and the outer energy gap at $k=0$
is $\left|m\right|+\sqrt{\Delta^{2}+4V^{2}}$, which is always positive.
For $s=-1$, the dispersions of the two inner bands are $E_{i}=-Dk^{2}\pm\sqrt{\left(\Lambda-\Gamma\right)^{2}+(\gamma k)^{2}\cos^{2}\Theta}$
($i=2$ for $-$, and 3 for $+$). The inner energy gap at $k=0$
is $\left|m\right|-\sqrt{\Delta^{2}+4V^{2}}$. Without $m$, the inner
bands and outer bands touch at $k=0$ [Fig. \ref{fig:evolution}
(a)]. A finite $m$ can lift the degeneracies at $k=0$ [Fig.
\ref{fig:evolution} (b)]. Increasing $m$ then produces a band
inversion for bands 2 and 3 at $k=0$, when
\begin{eqnarray}
|m|=\sqrt{\Delta^{2}+4V^{2}},\label{m-Delta-V}
\end{eqnarray}
and changes their topological properties from trivial to nontrivial
or vice versa [Figs. \ref{fig:evolution} (c) and \ref{fig:evolution}
(d)]. We have shown in Eq. (\ref{m-Delta-V}) that, in the presence
of SIA, the band inversion requires the exchange field $m$ to exceed
not $\Delta$ but a larger value $\sqrt{\Delta^{2}+4V^{2}}$.

\begin{figure}[htbp]
\centering \includegraphics[width=8.5cm]{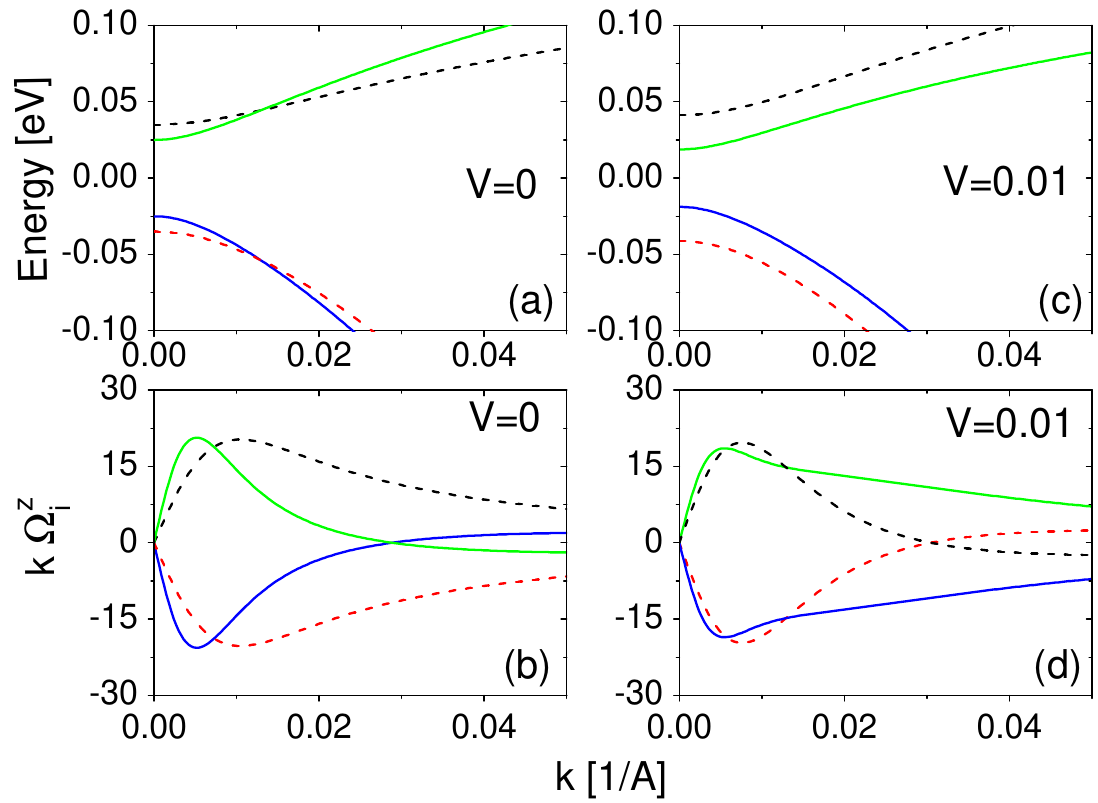} \caption{(color online). (a) There are band crossings if the band inversion
is achieved by magnetically doping a thin film with $\Delta B>0$ . (b)
The Berry curvature of the corresponding bands in (a). We have adopted
polar coordinates so $\Omega_{i}^{z}(\mathbf{k})\rightarrow k\Omega_{i}^{z}(k)$.
(c) The band crossings in (a) turn to anticrossings in the presence
of SIA ($V\neq0$). (d) The Berry curvature of the bands in (c).
Parameters: $\Delta=-0.01$ eV, $B=-30$ eV$\cdot$\AA$^{2}$, $\gamma=3$
eV$\cdot$\AA, $m=-0.06$ eV, $D=28$ eV$\cdot$\AA$^{2}$. Left: $V=0$. Right:
$V=0.01$ eV. The band with the lowest (highest) energy at $k=0$
is indexed as band 1 (4). Band 1: red dashed. Band 2: blue solid.
Band 3: green solid. Band 4: black dashed. }

\label{fig:SIA}
\end{figure}

Moreover, we find that SIA also leads to qualitative differences:
it makes the bands closest to the gap always topologically nontrivial.
Without SIA ($V=0$), the model reduces to the one proposed by
Yu \emph{et al} \cite{Yu10sci}. From the band structure, we can show
that if $\Delta B>0$, there will be band crossings
at a finite $k=\sqrt{\Delta/2B}$ [Fig. \ref{fig:evolution} (e)].
We find that SIA can turn the band
crossings at finite $k$ into band anticrossings, leading to a change
of topological properties {[}Fig. \ref{fig:evolution} (f){]}. As
a result, the inner two bands, 2 and 3, become nontrivial. To show
this, we investigate the topological properties of the bands with
the help of the Berry curvature
\begin{eqnarray*}
\Omega_{i}^{z}(\mathbf{k})=-2\sum_{j\neq i}\frac{\mathrm{Im}\langle i|\partial H/\partial k_{x}|j\rangle\langle j|\partial H/\partial k_{y}|i\rangle}{(E_{i}-E_{j})^{2}},
\end{eqnarray*}
where for given $\mathbf{k}$, $\left|i\right\rangle $ is the eigenstate in band $i$ with energy $E_{i}$. Usually, $\Omega_{i}^{z}(\mathbf{k})$ of
band $i$ is a function of $k$, and can have either positive or negative
values. Its integral over all the $k$ in the first Brillouin zone
is always an integer, i.e., the Chern number. If the Chern number
is zero, the energy band is topologically trivial. However, if $\Omega_{i}^{z}(\mathbf{k})$
is always positive or negative, its integral must be a nonzero integer,
meaning that the band is nontrivial. Figure \ref{fig:SIA} shows a
case in which the quantum anomalous Hall phase is created from a thin
film with $\Delta B>0$ and $m$ has already overcome $\sqrt{\Delta^{2}+4V^{2}}$
to invert bands 2 and 3 at $k=0$ [Fig. \ref{fig:SIA} (a)]. Without
SIA ($V=0$), the Berry curvature in Fig. \ref{fig:SIA} (b) shows
that bands 2 and 3 are trivial while bands 1 and 4 are nontrivial.
With SIA ($V\neq0$), the band crossings in Fig. \ref{fig:SIA} (a)
turn into anticrossings in Fig. \ref{fig:SIA} (c). Meanwhile, bands
2 and 3 exchanged their Berry curvature with bands 1 and 4 for $k>\sqrt{\Delta/2B}$ [Fig. \ref{fig:SIA} (d)].
As a result, bands 2 and 3 become nontrivial. Therefore, we have shown
that the inner bands are always nontrivial. In contrast, if $\Delta B<0$
there is no band crossing-anticrossing transition (see Fig. 1 of the Supplemental
Material \cite{supp}), and the inner two bands are topologically nontrivial
only because of the band inversion at $k=0$. Thus, in the presence
of SIA, the bands closest to the gap are always topologically nontrivial. We shall show that the nontrivial
bands account for the measured longitudinal and Hall conductances
in the following calculation.

\emph{Hall conductance}. - According to the linear response theory, the
intrinsic Hall conductance is given by the integral of the Berry curvature
over occupied states in $k$ space \cite{Xiao11rmp}
\begin{eqnarray*}
\sigma_{xy} & = & -\frac{e^{2}}{\hbar}\sum_{i}\int\frac{d^{2}\mathbf{k}}{(2\pi)^{2}}f(E_{i}-E_{F})\Omega_{i}^{z}(\mathbf{k}),
\end{eqnarray*}
where $f(x)=[\mathrm{exp}(x/k_{B}T)+1]^{-1}$ is the Fermi function
and $E_{F}$ is the Fermi energy. Without SIA, it has been shown that
\cite{Lu10prb} the in-gap Hall conductance $\sigma_{xy}^{\mathrm{gap}}=-\frac{e^{2}}{2h}[\mathrm{sgn}(\Delta+m)+\mathrm{sgn}(-\Delta+m)]$.
For the quantized anomalous Hall effect, $|m|$ must be larger than
$|\Delta|$, and then $\sigma_{xy}^{\mathrm{gap}}=-\frac{e^{2}}{h}\mathrm{sgn}(m)$,
which only depends on the sign of $m$. A positive quantized plateau
of $\sigma_{xy}$ indicates that $m<0$. In the presence of SIA, the value of the quantized Hall conductance remains the same but the quantum anomalous Hall phase
requires $|m|>\sqrt{\Delta^{2}+4V^{2}}$. In the experiment
\cite{Chang13sci}, the drop of the Hall conductance (inset of Fig.
\ref{fig:sigma}) from the quantized plateau is rather prompt on the
positive gate voltage side. This behavior implies that the band closest
to the gap is topologically nontrivial (band 3). The nonzero Berry
curvature of the band is mainly distributed near its band edge, leading
to the prompt drop of $\sigma_{xy}$. The concentrated Berry curvature
distribution is probably given by a relatively narrow bandwidth, and
small band gap. Another feature in $\sigma_{xy}$ is a small dip in
$\sigma_{xy}$ (marked by the arrow in Fig. \ref{fig:sigma}). This
is given by a trivial band at higher energy (band 4). If the band
4 is fully occupied, it has no contribution to the total Hall conductance.
But before being fully occupied, with increasing $E_{F}$, it first reduces
then enhances the Hall conductance, giving rise to the dip. By definition,
the intrinsic Hall conductance will always vanish at higher energy.
However, $\sigma_{xy}$, in the experiment, saturates at some finite
values, in particular, rather high ($\sim0.5e^{2}/h$) on the negative
gate voltage side. The pinning of the Fermi surface probably does not
account for the saturation, because $\sigma_{xx}$ is still increasing
meanwhile. The impurity scattering induced extrinsic Hall conductance
\cite{Nagaosa10rmp} may be one of the reasons. We calculated the
extrinsic contribution from the side-jump mechanism \cite{supp}.
However, it turns out that the perturbation theory assuming weak impurity
scattering \cite{Sinitsyn07prb,Yang11prb,Culcer11prb,Lu13arXiv} may not be enough, because the experiment was in a strong disordered or bad-metal regime \cite{Chang13sci,Chang13am}.

\begin{figure}[htbp]
\centering \includegraphics[width=8.5cm]{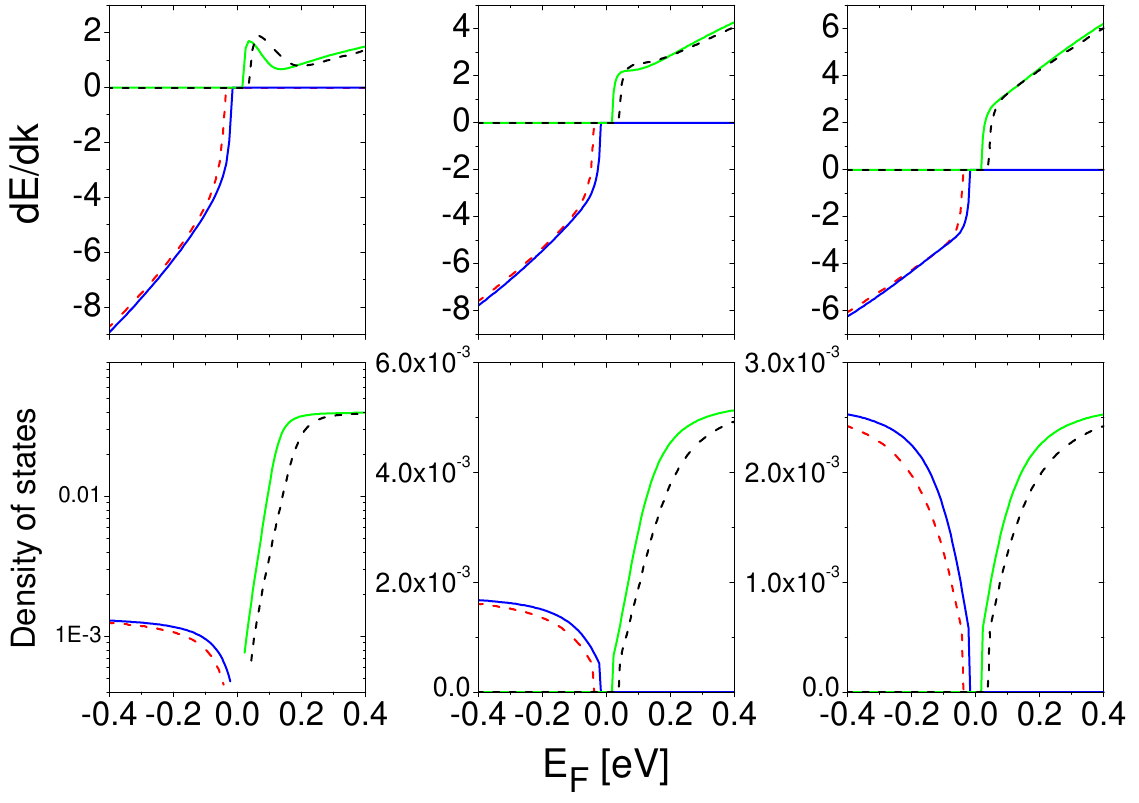} \caption{(color online). Group velocity $dE/dk$ and the density of states of bands
1-4. Parameters: $\Delta=0.01$, $B=-30$ eV$\cdot$\AA$^{2}$, $\gamma=3$
eV$\cdot$\AA, $m=-0.06$ eV, and $V=0.01$ eV. From left to right: $D=28$,
15, and 0 eV$\cdot$\AA$^{2}$. The colors and styles follow those in Fig.
\ref{fig:SIA}. The case with $\Delta=-0.01$ shows a similar result.}

\label{fig:group velocity}
\end{figure}

\emph{Longitudinal conductivity}. - Considering the low mobility in the
experimental sample, the electronic transport outside the gap is diffusive.
To the leading order, the longitudinal conductivity in the diffusive
regime is given by the Einstein relation \cite{Ashcroft-Mermin} $\sigma_{xx}=e^{2}N_{F}\mathcal{D}$,
where $N_{F}$ is the density of states at the Fermi energy, the diffusion
coefficient $\mathcal{D}=v_{F}^{2}\tau/2$, with the Fermi velocity
defined as $v_{F}=(\partial E/\partial k)/\hbar$, the scattering
time $\tau=\hbar/(2\pi N_{F}nu^{2})$ where $n$ is the impurity concentration,
$u$ measures the scattering strength. For simplicity, we use $nu^{2}$
to account for the overall effect from both the nonmagnetic and magnetic
scattering. Both $\sigma_{xx}$ and $\tau$ depend on $N_{F}$, indicating
that a large density of states also means strong scattering on the
Fermi surface. As a result, the density of states cancels out in $\sigma_{xx}$,
and we arrive at
\begin{eqnarray}
\sigma_{xx}=\frac{e^{2}}{h}\sum_{i=1}^{4}\frac{1}{2}\left|\frac{\partial E_{i}}{\partial k}\right|_{E_{i}=E_{F}}^{2}\frac{1}{nu_{i}^{2}}
\end{eqnarray}
for multiple bands. Suppose that $nu_{i}^{2}$ varies slowly with the Fermi energy; the peak in $\sigma_{xx}$, therefore, corresponds to a local maximum in
the group velocity. This behavior can be captured by the parameters
of the same orders as those in the experiments. In Fig. \ref{fig:group velocity},
we show the group velocity and the density of states {[}given by $k_{F}/(2\pi|dE/dk|_{k_{F}})$
in polar coordinates, $k_{F}$ the Fermi wave vector{]}. Compared
with Fig. \ref{fig:sigma}, it is clear that $\sigma_{xx}$ is dominantly
determined by the group velocity rather than the density of states.
It is found that the peak is better produced by comparable $|B|$
and $|D|$. Although Fig. \ref{fig:group velocity} shows only a case
with $\Delta B<0$, similar $\sigma_{xx}$ can be produced with $\Delta B>0$.

\emph{Thick films}. - For those thicker films (e.g., typically more than
5 nm for Bi$_{2}$Se$_{3}$), the hybridization between the top and
bottom surfaces becomes negligibly small, and both $\Delta$ and $B$
may be abandoned. In this case, the Hamiltonian can be projected to
the top and bottom surfaces \cite{supp}
\begin{eqnarray*}
U^{\dag}HU=-Dk^{2}\pm V+\frac{m}{2}\sigma_{z}+\gamma(k_{x}\sigma_{y}-k_{y}\sigma_{x}),
\end{eqnarray*}
where $U=(\sigma_{0}-i\sigma_{y})\otimes\sigma_{0}/\sqrt{2}$, and
$\pm V$ shows that SIA can shift the relative positions of the bands
from the top ($+$) surface with respect to those from the bottom ($-$) surface.
As massive Dirac fermion systems, each of the top and bottom surfaces
contributes $-(e^2/2h)\mathrm{sgn}(m)$ to the Hall conductance
when the Fermi level is located in the gap \cite{Redlich84prd,Jackiw84prd},
so the total Hall conductance is $-(e^2/h)\mathrm{sgn}(m)$.
For a smaller $\left|V\right|<\left|m\right|/2$, there is still an
energy gap that protects the quantized Hall conductance. However,
for a larger $V>|m|/2$, the energy gap closes and the Hall conductance
is not quantized. Since the SIA comes from the substrate-induced electric
field along the growth direction, a thicker film has a larger $V$ \cite{Zhang10natphys}.
Thus, if no other mechanism is taken into account, it may be hard to
realize a quantized anomalous Hall conductance in thicker films, although
the finite-size $\Delta$ is smaller. Also, in this case, the
surface states of the lateral sides of the film contribute to the
longitudinal conductance \cite{Chu11prb}, which forces the Hall resistance
to deviate from the quantized value.

We thank Michael Ma for helpful discussions on the longitudinal transport.
This work was supported by the Research Grant Council of Hong Kong
under Grant No. HKU7051/11P.


\begin{thebibliography}{10}
\bibitem{Nagaosa10rmp} N. Nagaosa, J. Sinova, S. Onoda, A. H. MacDonald, and N. P. Ong, Rev. Mod. Phys. \textbf{82}, 1539 (2010).

\bibitem{Xiao11rmp} D. Xiao, M. C. Chang, and Q. Niu, Rev. Mod. Phys.
\textbf{82}, 1959 (2010).

\bibitem{Haldane-88prl}F. D. M. Haldane, Phys. Rev. Lett. \textbf{61},
2015 (1988).

\bibitem{Liu08prl} C. X. Liu, X. L. Qi, X. Dai, Z. Fang, and S. C.
Zhang Phys. Rev. Lett. \textbf{101}, 146802 (2008).

\bibitem{Yu10sci} R. Yu, W. Zhang, H. J. Zhang, S. C. Zhang, X. Dai,
and Z. Fang, Science \textbf{329}, 61 (2010).

\bibitem{Qiao10rc} Z. Qiao, S. A. Yang, W. Feng, W. K. Tse, J. Ding,
Y. Yao, J. Wang, and Q. Niu, Phys. Rev. B \textbf{82}, 161414(R) (2010).

\bibitem{Qi06prb} X. L. Qi, Y. S. Wu, and S. C. Zhang, Phys. Rev.
B \textbf{74}, 085308 (2006).

\bibitem{Onoda06prl} S. Onoda, N. Sugimoto, and N. Nagaosa, Phys.
Rev. Lett. \textbf{97}, 126602 (2006).

\bibitem{Nomura11prl} K. Nomura and N. Nagaosa, Phys. Rev. Lett.
\textbf{106}, 166802 (2011).

\bibitem{Chu11prb} R. L. Chu, J. R. Shi, and S. Q. Shen, Phys. Rev.
B \textbf{84}, 085312 (2011).

\bibitem{Chang13am} C. Z. Chang, J. Zhang, M. Liu, Z. Zhang, X. Feng,
K. Li, L. L. Wang, X. Chen, X. Dai, Z. Fang, X. L. Qi, S. C. Zhang,
Y. Wang, K. He, X. C. Ma, and Q. K. Xue, Adv. Mater. \textbf{25}, 1065
(2013).

\bibitem{Checkelsky12natphys} J. G. Checkelsky, J. Ye, Y. Onose,
Y. Iwasa, and Y. Tokura, Nat. Phys. \textbf{8}, 729 (2012).

\bibitem{Hor10prb} Y. S. Hor, P. Roushan, H. Beidenkopf, J. Seo,
D. Qu, J. G. Checkelsky, L. A. Wray, D. Hsieh, Y. Xia, S.-Y. Xu, D.
Qian, M. Z. Hasan, N. P. Ong, A. Yazdani, and R. J. Cava, Phys. Rev.
B \textbf{81}, 195203 (2010).

\bibitem{Chen10sci} Y. L. Chen, J.-H. Chu, J. G. Analytis, Z. K.
Liu, K. Igarashi, H.-H. Kuo, X. L. Qi, S. K. Mo, R. G. Moore, D. H.
Lu, M. Hashimoto, T. Sasagawa, S. C. Zhang, I. R. Fisher, Z. Hussain, and
Z. X. Shen, Science \textbf{329}, 659 (2010).

\bibitem{Wray11natphys} L. A. Wray, S. Y. Xu, Y. Xia, D. Hsieh, A.
V. Fedorov, Y. S. Hor, R. J. Cava, A. Bansil, H. Lin, and M. Z. Hasan,
Nat. Phys. \textbf{7}, 32 (2011).

\bibitem{Chang13sci} C. Z. Chang, J. Zhang, X. Feng, J. Shen, Z.
Zhang, M. Guo, K. Li, Y. Ou, P. Wei, L. L. Wang, Z. Q. Ji, Y. Feng,
S. Ji, X. Chen, J. Jia, X. Dai, Z. Fang, S. C. Zhang, K. He, Y. Wang,
L. Lu, X. C. Ma, and Q. K. Xue, Science \textbf{340}, 167 (2013).

\bibitem{Zhang09np} H. J. Zhang, C. X. Liu, X. L. Qi, X. Dai, Z.
Fang, and S. C. Zhang, Nat. Phys. \textbf{5}, 438 (2009).

\bibitem{Shen11spin} S. Q. Shen, W. Y. Shan, and H. Z. Lu, SPIN \textbf{01},
33 (2011).

\bibitem{Lu10prb} H. Z. Lu, W. Y. Shan, W. Yao, Q. Niu, and S. Q.
Shen, Phys. Rev. B \textbf{81}, 115407 (2010).

\bibitem{Shan10njp} W. Y. Shan, H. Z. Lu, and S. Q. Shen, New J.
Phys. \textbf{12} 043048 (2010).


\bibitem{supp} See Supplementary Material for the calculation details.


\bibitem{Linder09prb} J. Linder, T. Yokoyama, and A. Sudb{\o},
Phys. Rev. B \textbf{80}, 205401 (2009).

\bibitem{Liu10rc} C. X. Liu, H. J. Zhang, B. Yan, X. L. Qi, T. Frauenheim,
X. Dai, Z. Fang, and S. C. Zhang, Phys. Rev. B \textbf{81}, 041307(R)
(2010).

\bibitem{Zhang10natphys} Y. Zhang, K. He, C. Z. Chang, C. L. Song,
L. L. Wang, X. Chen, J. F. Jia, Z. Fang, X. Dai, W. Y. Shan, S. Q.
Shen, Q. Niu, X. L. Qi, S. C. Zhang, X. C. Ma, and Q. K. Xue, Nat.
Phys. \textbf{6}, 584 (2010).

\bibitem{Sakamoto10prb} Y. Sakamoto, T. Hirahara, H. Miyazaki, S.
I. Kimura, and S. Hasegawa, Phys. Rev. B \textbf{81}, 165432 (2010).

\bibitem{Sinitsyn07prb} N. A. Sinitsyn, A. H. MacDonald, T. Jungwirth,
V. K. Dugaev, and J. Sinova, Phys. Rev. B \textbf{75}, 045315 (2007).

\bibitem{Yang11prb} S. A. Yang, H. Pan, Y. Yao, and Q. Niu, Phys.
Rev. B \textbf{83}, 125122 (2011).

\bibitem{Culcer11prb} D. Culcer and S. Das Sarma, Phys. Rev. B \textbf{83},
245441 (2011).

\bibitem{Lu13arXiv} H. Z. Lu and S. Q. Shen, Phys. Rev. B \textbf{88}, 081304(R) (2013).

\bibitem{Ashcroft-Mermin} N. W. Ashcroft and N. D. Mermin, \emph{Solid
State Physics} (Saunders, Philadelphia, 1976).

\bibitem{Redlich84prd} A. N. Redlich, Phys. Rev. D \textbf{29}, 2366
(1984).

\bibitem{Jackiw84prd} R. Jackiw, Phys. Rev. D \textbf{29}, 2375 (1984).

\end{thebibliography}
\end{document}